\documentclass{aa}
\usepackage{graphicx}
\usepackage{txfonts}
\begin{document}
   \title{The birth rate of subluminous and overluminous type Ia supernovae}


   \author{Xiangcun Meng
          \inst{1},
          Wencong Chen \inst{2,3},
          Wuming Yang\inst{1}
          and
          Zhongmu Li\inst{4,5}}

   \offprints{X. Meng}

   \institute{$^{\rm 1}$School of Physics and Chemistry, Henan Polytechnic
University, Jiaozuo, 454000, China, \email{xiangcunmeng@hotmail.com}\\
              $^{\rm 2}$Department of Physics, Shangqiu Normal University, Shangqiu, 476000,
              China\\
              $^{\rm 3}$Key Laboratory of Modern Astronomy and Astrophysics (Nanjing
University), Ministry of Education, Nanjing 210093,
China\\
              $^{\rm 4}$College of Physics and Electronic Information, Dali University, Dali, 671003,
              China\\
              $^{\rm 5}$National Astronomical Observatories, Chinese
Academy of Sciences, Beijing, 100012, China
             }

   \date{Received; accepted}


  \abstract
   {Based on the single degenerate (SD) scenario, a super-Chandrasekhar mass model derived from the rapid rotation of a progenitor star may
   account for the over-luminous type Ia supernovae (SNe Ia) like SN 2003fg. Chen \& Li (2009) calculated a series of binary evolution and showed
   the parameter spaces for the super-Chandrasekhar mass model.  Pakmor et al. (2010) developed a equal-mass
   double degenerate (DD) model to explain sub-luminous SNe Ia like SN 1991bg. But they did not
   show the evolution of the birth rate of these peculiar SNe Ia or did not compare them with absolute birth rates from observations.}
   {We want to show the evolution of the birth rates of these peculiar SNe Ia based on
   their results, and compare the birth rates with observations to check whether these model may account for all the peculiar SNe Ia.
}
   {We carried out a series of binary population synthesis
   calculations and considered two treatment of common envelope (CE) evolution,
   i.e. $\alpha$-formalism and $\gamma$-algorithm.
}
   {We found that the evolution of birth rate of these peculiar SNe Ia is heavily dependent
   on how to treat the CE evolution. The over-luminous SNe Ia may only occur
   for $\alpha$-formalism with low CE ejection efficiency and the delay time of the SNe Ia is
   between 0.4 and 0.8 Gyr. The upper limit of the contribution rate of the supernovae to
   all SN Ia is less than 0.3\%. The delay time of sub-luminous SNe Ia from equal-mass DD systems is
   between 0.1 and 0.3 Gyr for $\alpha$-formalism with $\alpha=3.0$, while longer than 9 Gyr for $\alpha=1.0$.
   The range of the delay time for $\gamma$-algorithm is very wide, i.e. longer
   than 0.22 Gyr, even as long as 15 Gyr. The sub-luminous SNe Ia from equal-mass DD systems
   may only account for no more than
   1\% of all SNe Ia observed.
   }
   {The super-Chandrasekhar mass model of Chen \& Li (2009) may account for a part of 2003fg-like
   supernovae and the equal-mass DD model developed by Pakmor et al. (2010) may explain some
   1991bg-like events, too. In addition, based on the comparison between theories and observations,
   including the birth rate and delay time of the 1991bg-like events,
   we found that the $\gamma$-algorithm is more likely to be
   an appropriate prescription of
   the CE evolution of DD systems than the $\alpha$-formalism if the equal-mass DD
   systems is the progenitor of 1991bg-like SNe Ia.
   }

   \keywords{Stars: white dwarfs - stars: supernova: general -  supernovae: individual:
   SN 2003fg, SN 1991bg
               }
   \authorrunning{Meng, Chen, Yang \& Li}
   \titlerunning{The birth rate of subluminous and overluminous type Ia supernova}
   \maketitle{}
%

\section{Introduction}\label{sect:1}
As one of distance indicators, type Ia supernovae (SNe Ia) showed
their importance in determining cosmological parameters, which
results in the discovery of the accelerating expansion of the
universe (Riess et al. \cite{RIE98}; Perlmutter et al.
\cite{PER99}). The result is exciting and suggesting the presence
of dark energy. At present, SNe Ia are proposed to be cosmological
probes for testing the evolution of the dark energy equation of
state with time and testing the evolutionary history of the
universe (Riess et al. \cite{RIESS07}; Kuznetsova et al.
\cite{KUZNETSOVA08}; Howell et al. \cite{HOWEL09}). They even were
chose to check the general relativity (Zhao et
al.\cite{ZHAOGB10}). However, the nature of SNe Ia is still
unclear, especially its progenitor system (Hillebrandt \& Niemeyer
\cite{HN00}; Leibundgut \cite{LEI00}). It is widely believed that
a SN Ia is produced from the thermonuclear runaway of a
carbon-oxygen white dwarf (CO WD) in a binary system (Arnett
\cite{ARN82}). Based on the nature of the WD companion, two basic
scenarios have been discussed for about three decades. One is the
single degenerate (SD) model (Whelan \& Iben \cite{WI73}; Nomoto
et al. \cite{NTY84}), i.e. the companion may be a main-sequence
(MS) or a slightly evolved star (WD+MS) or a red giant star
(WD+RG) or a helium  star (WD + He star), and the model is widely
investigated by many groups (Yungelson et al. \cite{YUN95}; Li et
al. \cite{LI97}; Hachisu et al. \cite{HAC99a}, \cite{HAC99b};
Nomoto et al. \cite{NOM99, NOM03}; Langer et al. \cite{LAN00}; Han
\& Podsiadlowski \cite{HAN04}; Chen \& Li \cite{CHENWC07}; Han
\cite{HAN08}; Meng \& Yang \cite{MENGXC10b}; L\"{u} \cite{LGL09};
Wang et al. \cite{WANGB09a,WANGB09b}; Wang, Li \& Han
\cite{WANGB10}), the other is the double degenerate (DD) model
(Iben \& Tutukov \cite{IBE84}; Webbink \cite{WEB84}), in which a
system of two CO WDs loses orbital angular momentum by means of
gravitational wave radiation and finally merges. Although the
birth rate and the distribution of its delay time (DDT, delay
time: between the episode of star formation producing progenitor
systems and the occurrence of SNe Ia) from this channel is
comparable to observations (Han \cite{HAN98}; Yungelson \& Livio
\cite{YUN98,YUN00}; Tutukov \& Yungelson \cite{TUT02}; Mennekens
et al. \cite{Mennekens10}) and some SNe Ia such as SN 2003fg and
SN 2005hj may likely result from the channel (Howell et al.
\cite{HOW06}; Quimby et al. \cite{QUI07}), some numerical
simulations showed that the most probable fate of the coalescence
is an accretion-induced collapse and, finally, neutron star
formation (see the review by Hillebrandt \& Niemeyer \cite{HN00}).
Generally, the WD mass at the moment of explosion for SD model is
equal (the so-called Chandrasekhar mass model) or less than the
Chandrasekhar limit (the sub-Chandrasekhar mass model), while
larger than the limit for DD model (super-Chandrasekhar mass
model).

The most remarkable property of SNe Ia is their apparent
homogeneour nature. This characteristic is interpreted as the
observable consequence of a model where the progenitors of SNe Ia
are CO WDs which have increase their mass to close to the
Chandrasekhar limit via mass transfer in a binary system.
Nevertheless, observations of two peculiar events, i.e. SN 1991T
and SN 1991bg, raised questions about the uniformity of SNe Ia.
Motivated by these discoveries, Phillips (\cite{PHI93}) suggested
a relation between the absolute magnitude of SNe Ia at maximum
light and the magnitude drop of B light curve during the first 15
days following maximum.  When SNe Ia are applied as distance
indicator, the relation is adopted.

The peak brightness of SN 1991bg was $\sim2.0$ mag fainter in B
than that of normal SNe Ia, and its light curve declined unusually
fast following maximum (Filippenko et al. \cite{FILIPPENKO92b};
Leibundgut et al. \cite{LEIBUNDGUT93}). SN 1991bg is identified by
Branch \& Miler (\cite{BM93}) as intrinsically subluminous. The
low peak luminosity indicated a low amount of $^{\rm 56}$Ni, i.e.
about 0.07 $M_{\odot}$ (Mazzali et al. \cite{MAZZALI97}).
Generally, the subluminous property of the SN 1991bg may be
explained by a sub-Chandrasekhar mass model from a CO WD + He star
system or a merger of CO WD and He WD (Branch et al. \cite{BRA95};
Livio \cite{LIV99,LIV03}). However, Pakmor et al.
(\cite{PAKMOR10}) recently suggested an alternative scenario, i.e.
sub-luminous 1991bg-like events may be from the mergers of
equal-mass CO WDs of $M\sim0.9M_{\odot}$. Although the light curve
from the merger model is broader than that of 1991bg-like events,
the synthesized spectra, red color and low expansion velocities
are all close to those observed for 1991bg-like events. They
claimed that the events from the merger of equal-mass WDs should
occur with a rate of $\approx2-11\%$ of all SNe Ia. However,
Pakmor et al. (\cite{PAKMOR10}) discussed a relative birth rate to
various SNe Ia progenitor channels rather than the absolute ones
from observations. One of the purposes in this paper is to present
the evolution of the birth rate of the sub-luminous 1991bg-like
events from the mergers of equal-mass WDs and to compare the birth
rate with those from observations.

SN 1991T is an overluminous event and its light curve declined
slowly after maximum luminosity (Filippenko et al.
\cite{FILIPPENKO92a}; Phillips et al. \cite{PHILLIPS92,PHI99}),
which is often taken as an indication of a large $^{\rm 56}$Ni
mass (H\"{o}flich et al. \cite{HOF95}; Nugent et al. \cite{NUG97};
Pinto \& Eastman \cite{PE01}). Nevertheless, Kasen et al.
(\cite{KAS04}) suggested a second, physically very different route
to explain the peculiarities of SN 1991T--- one could be peering
down an ejecta hole which is due to the existence of the companion
at the moment of supernova explosion. This suggestion was uphold
by detailed binary population synthesis (BPS) study (Meng, \& Yang
\cite{MENGXC10a}). So, it is possible that 1991T-like supernovae
have not any special properties in physics except for the viewing
angle of an observer.

The SN 2003fg was observed to be 2.2 times over-luminous than a
normal SN Ia and the amount of $^{\rm 56}$Ni was inferred to be
$1.3M_{\odot}$, which requires a super-Chandrasekhar mass
explosion ($\sim2.1 M_{\odot}$, Astier et al. \cite{ASTIER06};
Howell et al. \cite{HOW06}). Since SN 2003fg, another three
2003fg-like events were discovered, i.e. SN 2006gz (Hicken et al.
\cite{HICKEN07}), SN 2007if (Scalzo et al. \cite{SCALZO10}; Yuan
et al. \cite{YUAN10}) and SN 2009dc (Tanaka et al.
\cite{TANAKA10}; Yamanaka et al. \cite{YAMANAKA10}). These
supernovae are usually assumed from the mergers of DD systems,
where the total mass of the DD systems is over the Chandrasekhar
mass limit. However, a super-Chandrasekhar WD may also exist in a
SD system, where the massive WD is supported by rapid rotation
(Yoon \& Langer \cite{YOON05}; Howell et al. \cite{HOW06}). For
example, Maeda \& Iwamoto (\cite{MAEDA09}) showed that the
properties of SN 2003fg are more consistent with aspherical
explosion of a super-Chandrasekhar WD supporting by a rapid
rotation. Taking account of the influence of rotation on accreting
WD, Yoon \& Langer (\cite{YOON04}) found that a WD under a special
condition may rotate differentially. Using the result of Yoon \&
Langer (\cite{YOON04}), Chen \& Li (\cite{CHENWC09}) calculated
the evolution of close binaries consisting of a CO white dwarf and
a normal companion, and obtained the parameter space in an orbital
period - secondary mass ($\log P_{\rm i}, M_{\rm 2}^{\rm i}$)
plane for super-Chandrasekhar SNe Ia. However, they also did not
present the evolution of the birth rate of the super-Chandrasekhar
SNe Ia. In this paper, we will show the evolution by a detailed
BPS study.

In Sect. \ref{sect:2}, we describe our BPS method. We show the BPS
results in Sect. \ref{sect:3}, and the discussions and conclusions
in Sect. \ref{sect:4}.


\section{Binary population synthesis}\label{sect:2}

\subsection{Super-Chandrasekhar SNe Ia}\label{sect:2.1}
The study of Chen \& Li (\cite{CHENWC09}) is based on a WD + MS
system and their results were summarized in an orbital period -
secondary mass ($\log P_{\rm i}, M_{\rm 2}^{\rm i}$) plane.
According to the situation of the primary in a primordial system
at the onset of the first Roche lobe overflow (RLOF), there are
three evolutionary channels contributing to the WD + MS system,
i.e. He star channel, EAGB channel and TPAGB channel (see Meng,
Chen \& Han \cite{MENGXC09} and Meng \& Yang \cite{MENGXC10b} for
details about the channels). In this paper, the TPAGB channel is
the dominant one since the initial mass of WD for over-luminous
SNe Ia is larger than 1.0 $M_{\odot}$, and the other two channels
do not contribute to the over-luminous because the initial WD mass
from this channel is always smaller than 1.0 $M_{\odot}$ (see also
Fig. 8 in Meng, Chen \& Han \cite{MENGXC09}). For all the three
channels, a common envelope (CE) phase is expected. After the
formation of the WD + MS systems, the systems continue to evolve
and the secondaries may also fill their Roche lobes at a stage and
RLOF starts. We assume that if the initial orbital period, $P_{\rm
orb}^{\rm i}$, and the initial secondary mass, $M_{\rm 2}^{\rm
i}$, of a SD system locate in the appropriate regions in the
($\log P^{\rm i}, M_{\rm 2}^{\rm i}$) plane for
super-Chandrasekhar SNe Ia at the onset of RLOF, a
super-Chandrasekhar SN Ia is then produced.

\subsection{Sub-Luminous SNe Ia}\label{sect:2.2}
Pakmor et al. (\cite{PAKMOR10}) also made a BPS study about
1991bg-like events based on their merging model from equal-mass DD
systems, where the components are both massive and the mass ratio
is only slightly less than one, i.e. the primary masses are in the
range of $0.85-1.05$ $M_{\odot}$ with a mass ratio of $0.9<q<1.0$.
Since the results from the large WD mass (as large as 1.05
$M_{\odot}$) will more likely lead to more luminous SNe Ia, not
1991bg-like events, we critically limit that the range of WD mass
is between 0.83 and 0.9 $M_{\odot}$ in this paper (see also Pakmor
et al. \cite{PAKMOR10}). A primordial binary system may become a
DD system after one or two CE phases (Han \cite{HAN98}; Ruiter et
al. \cite{RUITER09}). There are three sub-categories to form the
DD system contributing to the sub-luminous SNe Ia. In the
following, we briefly outline the three sub-categories: Case 1, 2
and 3.

Case 1 (2RLOF + 2CE): Although the evolutionary channel for Case 1
is similar, the parameters of the primordial binaries for
different CE treatment are different (see next subsection). For
$\alpha$-formalism, the primordial zero-age main sequence (ZAMS)
mass of primaries is in the range from 5.2 $M_{\odot}$ to 5.7
$M_{\odot}$, and the mass ratio ($m_{\rm 2}/m_{\rm 1}$) is between
0.59 and 0.7, and the primordial separation is close, i.e. from 30
$R_{\odot}$ to 60 $R_{\odot}$. For $\gamma$-algorithm, the
primordial ZAMS mass of primaries is in the range from $\sim5.35$
$M_{\odot}$ to 7.1 $M_{\odot}$, and the mass ratio is between 0.35
and 0.53, and the primordial separation is even more close, i.e.
from 17 $R_{\odot}$ to 30 $R_{\odot}$. Due to the close
separation, the primary fills its Roche Lobe when it is crossing
Hertzprung Gap (HG), and then a stable RLOF occurs. The primary
loses its hydrogen-rich envelope and then becomes a helium star,
where mass ratio reverses at a point. The helium star may fill its
Roche lobe again after its central helium is exhausted, where the
RLOF is also dynamically stable for low mass ratio. During the two
stable RLOF, the separation may increase greatly for material
transferring from light component to heavy one, and then the
system finally becomes a very wide WD + MS system, where the
separation may be as large as 400 $R_{\odot}$. The secondary fills
its Roche lobe during HG, which leads to a CE for a high mass
ratio ($\sim10$). After the CE ejection, the mass donor becomes a
helium star and continues to evolve. The helium star fills its
Roche lobe once again after the exhaustion of the central helium,
and a second CE ensues (mass ratio $\sim2$). Thus a DD system
forms. However, the final separation is heavily dependent on the
treatment of the CE, i.e. $\leq0.6 R_{\odot}$ for the
$\alpha$-formalism, while $2-3$ $R_{\odot}$ for the
$\gamma$-algorithm, which is the dominant factor resulting in
different delay time of SNe Ia for the different treatment of the
CE.

Case 2 (RLOF + CE): For the case, the mass ratio for primordial
binary system is very close to 1, i.e. larger than 0.94, and the
primordial ZAMS mass of the primary is between 3.75 and 4.1
$M_{\odot}$. The primordial system has a very wide separation
(wider than 2500 $R_{\odot}$), which permits the primary to evolve
to thermal pulsing asymptotic giant branch (TPAGB) stage before it
fills its Roche lobe, and then a stable RLOF occurs. At this
stage, the secondary is a horizontal branch (HB) star (central
helium burning). The system consists of a CO WD and a HB star
after the RLOF. The secondary fills its Roche lobe again at TPAGB
stage and a CE forms for a high mass ratio($\sim7$). After the CE
ejection, the orbit decays greatly and a DD system forms with a
separation of 0.2 to 2 $R_{\odot}$. Then, the delay time of the
sub-luminous SNe Ia from this sub-category is less than 2 Gyr.

Case 3 (one CE): The primordial ZAMS mass of primaries is around
$4 M_{\odot}$ and the mass ratio is almost 1, i.e. larger than
0.999. The primordial separation is from $1200 R_{\odot}$ to $1600
R_{\odot}$, and thus both primary and secondary are at TPAGB stage
when primary fills its Roche lobe, which results in a CE. After
the CE ejection, a DD system forms with a separation of $\sim 3
R_{\odot}$. The delay time of SNe Ia from this sub-category is
longer than 9 Gyr.

Following the DD system, gravitational wave radiation (GW)
dominates the evolution of system on a timescale $t_{\rm GW}$
(Landau \& Lifshitz \cite{LANDAU62}),
\begin{equation}
t_{\rm GW}{\rm (yr)}= 8\times10^{\rm 7}\times\frac{(M_{\rm
1}+M_{\rm 2})^{\rm 1/3}}{M_{\rm 1}M_{\rm 2}}P^{8/3},\label{eq:gw}
  \end{equation}
where $P$ is the orbital period of the DD system in hours, and
$M_{\rm 1}$ and $M_{\rm 2}$ are the mass of the two white dwarf in
solar mass, respectively. Then, the time elapsed from the birth of
primordial binary system to the occurrence of SN Ia is equal to
the sum of the timescale on which the secondary star becomes a WD
and the orbital decay time. We assume that if $M_{\rm 1}$ and
$M_{\rm 2}$ are both in the range of $0.83-0.9$ $M_{\odot}$ and
the elapsed time is less than $15$ Gyr, a sub-luminous SN Ia is
produced.

\subsection{Common envelope}\label{sect:2.3}
As mentioned above, CE is very important for the formation of SD
and DD systems. During binary evolution, the mass ratio ($q=M_{\rm
donor}/M_{\rm accretor}$) is crucial. If it is larger than a
critical mass ratio, $q_{\rm c}$, mass transfer between the two
components is dynamically unstable and a CE forms
(Paczy$\acute{\rm n}$ski \cite{PAC76}). The ratio $q_{\rm c}$
varies with the evolutionary state of the donor star at the onset
of RLOF (Hjellming \& Webbink\cite{HW87}; Webbink
\cite{WEBBINK88}; Han et al. \cite{HAN02}; Podsiadlowski et al.
\cite{POD02}; Chen \& Han \cite{CHE08}). In this study, we adopt
$q_{\rm c}$ = 4.0 when the donor star is on MS or crossing HG.
This value is supported by detailed binary evolution studies (Han
et al. \cite{HAN00}; Chen \& Han \cite{CHE02, CHE03}). If the
primordial primary is on first giant branch (FGB) or asymptotic
giant branch (AGB), we use
\begin{equation} q_{\rm c}=[1.67-x+2(\frac{M_{\rm c}}{M})^{\rm 5}]/2.13,  \label{eq:qc}
  \end{equation}
where $M_{\rm c}$ is the core mass of the donor star, and $x={\rm
d}\ln R_{\rm 1}/{\rm d}\ln M$ is the mass--radius exponent of the
donor star and varies with composition. If the mass donors
(primaries) are naked helium giants, $q_{\rm c}$ = 0.748 based on
Equation (\ref{eq:qc}) (see Hurley et al. \cite{HUR02} for
details).

Embedded in the CE are the dense core of the donor star and the
secondary. Due to frictional drag with the envelope, the orbit of
the embedded binary decays, and a large part of the orbital energy
released in the spiral-in process is injected into the envelope
(Livio \& Soker \cite{LS88}). It is generally assumed that the CE
is ejected if
\begin{equation}
\alpha_{\rm CE}\Delta E_{\rm orb}\geq |E_{\rm bind}|,
\label{eq:alpha}
  \end{equation}
where $\Delta E_{\rm orb}$ is the orbital energy released, $E_{\rm
bind}$ is the binding energy of CE, and $\alpha_{\rm CE}$ is CE
ejection efficiency, i.e. the fraction of the released orbital
energy used to eject the CE. Since the thermal energy in the
envelope is not incorporated into the binding energy, $\alpha_{\rm
CE}$ may be greater than 1 (see Han et al. \cite{HAN95} for
details about the thermal energy). In this paper, we set
$\alpha_{\rm CE}$ to 1.0 or 3.0\footnote{Via reconstructing the
evolution of post-common-envelope binaries (PCEB) consisting of a
white dwarf and a main-sequence star, Zorotovic et al.
(\cite{ZOROTOVIC10}) obtained a relative low $\alpha_{\rm CE}$
value, where the internal energy of the envelope was included.}.
After the CE, the orbital separation at the onset of the CE,
$a_{\rm i}$, becomes $a_{\rm f}$ which is determined by
\begin{equation}
\frac{G(M_{\rm c}+M_{\rm e})M_{\rm e}}{\lambda R_{\rm
1}}=\alpha_{\rm CE}(\frac{GM_{\rm c}m}{2a_{\rm
f}}-\frac{GMm}{2a_{\rm i}}),
  \end{equation}
where $\lambda$ is a structure parameter relaying on the
evolutionary stage of the donor, $M$, $M_{\rm c}$ and $M_{\rm e}$
are the masses of the donor, the donor envelope and the core,
respectively, $R_{\rm 1}$ is the radius of the donor, and $m$ is
the companion mass. In the paper, we take the structure parameter
a constant, i.e. $\lambda=0.5$ (de Kool, van den Heuvel \& Pylyser
\cite{DEKOOL87}). Thus, the final orbital separation of a binary
system after the CE phase $a_{\rm f}$ is given by

\begin{equation}
\frac{a_{\rm f}}{a_{\rm i}}=\frac{M_{\rm c}}{M}(1+\frac{2M_{\rm
e}a_{\rm i}}{\alpha_{\rm CE}\lambda mR_{\rm 1}})^{\rm -1}.
  \end{equation}

The description above may well produce the distribution of WD + MS
systems as noticed by Zorotovic et al. (\cite{ZOROTOVIC10}) (see
also Hurley et al. \cite{HUR02}), while it is possibly difficult
for the description to produce a close pair of white dwarfs.
Nelemans et al. (\cite{NELEMANS00}) and Nelemans \& Tout
(\cite{NELEMANS05}) suggested an alternative algorithm equating
the angular momentum balance, which may explain the formation of
all kinds of close binaries:
\begin{equation}
\frac{\Delta J}{J}=\gamma_{\rm CE}\frac{M_{\rm
e}}{M+m},\label{eq:gamma}
  \end{equation}
where $J$ is the total angular momentum and $\Delta J$ is the
change of the total angular momentum during the common envelope
phase. Implicitly assuming the conservation of energy, the final
orbital separation $a_{\rm f}$ after the CE is then given by

\begin{equation}
\frac{a_{\rm f}}{a_{\rm i}}=(\frac{M}{M_{\rm c}})^{\rm
2}(\frac{M_{\rm c}+m}{M+m})(1-\gamma_{\rm CE}\frac{M_{\rm
e}}{M+m})^{\rm 2}.
  \end{equation}
Based on the results in Nelemans \& Tout (\cite{NELEMANS05}), we
set $\gamma_{\rm CE}=1.5$.

Following Nelemans \& Tout (\cite{NELEMANS05}), we call the the
formalism for equation (\ref{eq:alpha}) the $\alpha$-formalism and
that of equation (\ref{eq:gamma}) the $\gamma$-algorithm. In this
paper, we treat CE evolution by the two algorithms.

\subsection{Basic parameters for Monte Carlo simulations}\label{sect:2.4}
To investigate the birthrate of SNe Ia, we followed the evolution
of $4\times10^{\rm 7}$ binaries via Hurley's rapid binary
evolution code (Hurley et al. \cite{HUR00, HUR02}). The
descriptions above for super-luminous SNe Ia from the SD model and
the sub-luminous ones from the equal-mass DD systems are
incorporated into the code. The primordial binary samples are
generated in a Monte Carlo way and a circular orbit is assumed for
all binaries. The basic parameters for the simulations are as
follows: (1) a constant star formation rate (SFR) of $5$
$M_{\odot}$ ${\rm yr^{-1}}$ over the past 15 Gyr or a single star
burst of $10^{\rm 11} M_{\odot}$; (2) the initial mass function
(IMF) of Miller \& Scalo (\cite{MIL79}); (3) the mass-ratio
distribution is constant; (4)all stars are members of binary
systems and the distribution of separations is constant in $\log
a$ for wide binaries, where $a$ is the orbital separation, and
falls off smoothly at small separation, where $a=10R_{\odot}$ is
the boundary for wide and close binaries; (5) solar metallicity,
i.e. $Z=0.02$ (see Meng \& Yang (\cite{MENGXC10b}) for details of
the parameter input).

   \begin{figure}
   \centering
   \includegraphics[width=60mm,height=80mm,angle=270.0]{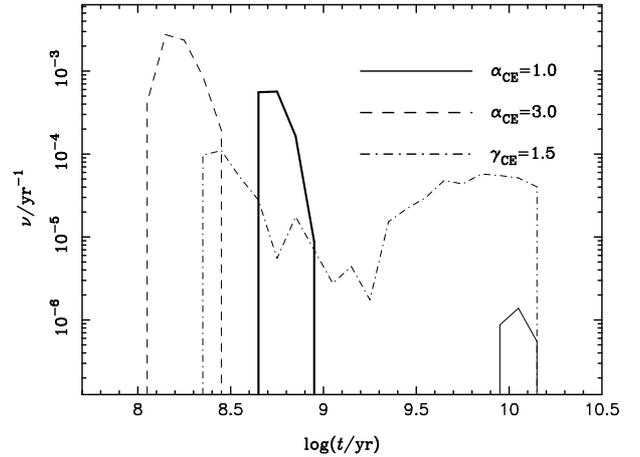}
   \caption{Evolution of the birthrates of super-luminous SNe Ia from SD channel and sub-luminous SNe Ia
   from the merge of equal-mass DD systems for a single
   starburst of $10^{\rm 11}M_{\odot}$ for different CE treatments.
   The thick line is the result for over-luminous SNe Ia,
   while the thin lines are the results for sub-luminous SNe Ia}
              \label{Fig1}%
    \end{figure}

   \begin{figure}
   \centering
   \includegraphics[width=60mm,height=80mm,angle=270.0]{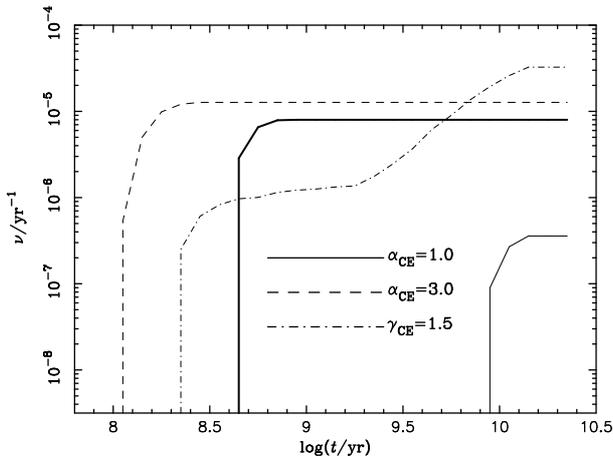}
   \caption{Similar to Fig. \ref{Fig1} but for a constant star formation
   rate of 5 $M_{\rm \odot}{\rm yr^{\rm -1}}$.} \label{Fig2}%
    \end{figure}

\section{Results}\label{sect:3}

\begin{table*}[] \caption[]{The Galactic birth rate for super- (Row 3) and sub-luminous (Row 4)
SNe Ia with different BPS parameters (Row 2), where a constant
star formation rate ($5$ $M_{\odot}$ ${\rm yr^{-1}}$) is assumed
over the past 15 Gyr.} \label{Tab:1}
\begin{center}
\begin{tabular}{cccc}
\hline\noalign{\smallskip}
Series Number &parameters &$\nu_{\rm 1}(10^{\rm -6}{\rm yr^{\rm -1}})$ &$\nu_{\rm 2}(10^{\rm -6}{\rm yr^{\rm -1}})$ \\
\hline\noalign{\smallskip}
1 & $\alpha_{\rm CE}=1.0$; $n(q)=1$; $e=0$  &7.98  &0.359   \\
2 & $\alpha_{\rm CE}=3.0$; $n(q)=1$; $e=0$  &0.0   &12.7   \\
3 & $\gamma_{\rm CE}=1.5$; $n(q)=1$; $e=0$  &0.0   &32.6   \\
4 & $\alpha_{\rm CE}=1.0$; $n(q)=2q$; $e=0$ &6.19  &0.807   \\
5 & $\alpha_{\rm CE}=1.0$; $n(q)=1$; $n(e)=2e$  &8.24  &7.88   \\
5 & $\alpha_{\rm CE}=2.0$; $n(q)=1$; $n(e)=2e$  &0.0  &1.97   \\
\hline\noalign{\smallskip} 
\end{tabular}\end{center}
\end{table*}

\subsection{Single star burst}\label{sect:3.1}
Fig. \ref{Fig1} shows the evolution of the birth rate of
over-luminous (the thick line) and sub-luminous (the thin lines)
SNe Ia with different CE treatments, where a single star burst is
assumed. An imposing property in the figure is that the results
are heavily dependent on the treatment of CE evolution. For
over-luminous SNe Ia, only $\alpha$-formalism with $\alpha=1.0$
may produce SNe Ia. This is mainly derived from constraint of the
parameter space for over-luminous SNe Ia and the different
evolutionary channels for WD + MS systems. Based on the results in
Chen \& Li (\cite{CHENWC09}), a CO WD may explode as an
over-luminous SN Ia only when its initial mass is larger than 1.0
$M_{\odot}$. As described in section \ref{sect:2.1}, there are
three channel may produce WD + MS system, but the TPAGB channel is
the only one leading to the WD + MS system with an initial WD mass
larger than 1.0 $M_{\odot}$. The primordial binary system
experiencing the TPAGB channel has a very long orbital period,
i.e. longer than 1000 days, which means a low binding energy of CE
(Meng, Chen \& Han \cite{MENGXC09}). Because of the low binding
energy of the CE and a long primordial orbital period,
$\alpha_{\rm CE}$ has a remarkable influence on CO+WD systems from
the TPAGB channel. Generally, if a CE can be ejected, a low
$\alpha_{\rm CE}$ produces a shorter orbital-period WD + MS
system, which is more likely to fulfill the conditions for
over-luminous SNe Ia. Therefore, we only obtain over-luminous SNe
Ia when $\alpha_{\rm CE}=1.0$ (see also Meng, Chen \& Han
\cite{MENGXC09}). For WD + MS systems, the effect of the
$\gamma$-algorithm is similar to that of the $\alpha$-formalism
with $\alpha_{\rm CE}=3.0$ (Nelemans \& Tout \cite{NELEMANS05}).
Thus no over-luminous SNe Ia is produced for the
$\gamma$-algorithm.

It is clearly shown in Fig. \ref{Fig1} that most of over-luminous
SNe Ia occur between 0.4 and 0.8 Gyr. This is mainly determined by
the high initial companion mass of WD (see Figs. 2 and 3 in Chen
\& Li \cite{CHENWC09}). In addition, the peak value of the birth
rate for over-luminous SNe Ia is much smaller than that for normal
SNe Ia from WD + MS systems by about two orders of magnitude (see
Han \& Podsiadlowski \cite{HAN04}; Meng, Chen \& Han
\cite{MENGXC09}).

Both $\alpha$-formalism and $\gamma$-algorithm may produce the
sub-luminous SNe Ia. However, the sub-luminous SNe Ia may be from
different sub-categories for different CE treatment. For
$\alpha$-formalism, sub-luminous SNe Ia are produced from Case 1
only when $\alpha_{\rm CE}$ has a high value, i.e. $\alpha_{\rm
CE}=3.0$. Because a binary system have experienced two CE phase
for Case 1 before it become a DD system suitable for sub-luminous
SNe Ia, a low $\alpha_{\rm CE}$ means a shorter orbital period,
even a merger before CE ejection, and then no sub-luminous SNe Ia.
Actually, when $\alpha_{\rm CE}=1.0$, all the binary system
suitable for sub-luminous SNe Ia merger before a DD system forms,
except those from Case 3. Since the primordial binary systems from
Case 3 have a very large separation, its final separation is also
very large after a CE phase. Thus, the delay time from this
channel is very long, i.e. longer than 9 Gyr. Whatever, the birth
rate from this channel is too low, i.e. the peak value is only
$10^{\rm -6}$ ${\rm yr^{\rm -1}}$ (see the thin solid line in Fig.
\ref{Fig1}). Even for $\alpha_{\rm CE}=3.0$, the result of
sub-luminous SNe Ia significantly differ from that for
$\gamma$-algorithm. The sub-luminous SNe Ia occur between 0.1 and
0.3 Gyr for $\alpha=3.0$, while explode with a very wide age range
for $\gamma$-algorithm, i.e. longer than 0.22 Gyr and even as long
as 15 Gyr. The sub-luminous SNe Ia for $\gamma$-algorithm come
from two sub-categories, i.e. Case 2 with delay time shorter than
2 Gyr, while Case 1 with delay time longer than 2 Gyr.

Another imposing property in Fig. \ref{Fig1} is that the peak
value of the birth rate is much lower than those from observations
by at least two orders of magnitude (Totani  et al.
\cite{TOTANI08}; Maoz et al. \cite{MAOZ10a}; Maoz \& Badenes
\cite{MAOZ10b}), and the shape of the DDT in this figure also
significantly differs from observations (Maoz et al.
\cite{MAOZ10c}). The best fitted DDT from observations follows
$t^{\rm -1}$ (Totani et al. \cite{TOTANI08}; Maoz et al.
\cite{MAOZ10c}), while the DDT is a narrow peak for the
super-luminous SNe Ia from the SD channel and the sub-luminous
ones from the merge of equal-mass DD systems under
$\alpha$-formalism. Although the DDT for the sub-luminous under
the $\gamma$-algorithm is not a narrow peak, its shape still does
not follow $t^{\rm -1}$. These results are good evidence for
sub-group nature of 2003fg-like and 1991bg-like SNe Ia.

\subsection{Constant star formation rate}\label{sect:3.2}
Fig. \ref{Fig2} shows Galactic birthrates of SNe Ia for
over-luminous and sub-luminous SNe Ia. Our simulations give an
upper limit of the birth rate of over-luminous SNe Ia from WD + MS
systems, i.e. $\sim8\times10^{\rm -6}$ yr$^{\rm -1}$ (thick solid
line), which is smaller than the Galactic birthrate inferred
observationally by about three orders of magnitude
(3-4$\times10^{\rm -3}{\rm yr^{\rm -1}}$, van den Bergh \& Tammann
\cite{VAN91}; Cappellaro \& Turatto \cite{CT97}; Li et al.
\cite{LIWD10}). So, the over-luminous SNe Ia from WD + MS systems
could be very rare events. The upper limit of the birth rate of
sub-luminous SNe Ia from the equal-mass DD systems is only
slightly higher than that of over-luminous SNe Ia, i.e.
$\sim3\times10^{\rm -5}{\rm yr^{\rm -1}}$, and then the
sub-luminous SNe Ia from DD systems are also rare events.

\section{Discussions and conclusions}\label{sect:4}
\subsection{Birth rate}\label{sect:4.1}
According to the results of Chen \& Li (\cite{CHENWC09}), we found
that the 2003fg-like supernovae from SD systems contribute to no
more than 0.3\% of all SNe Ia. The contribution rate is so low and
seems not to account for the discovery of four 2003fg-like events
at present. Considering the delay time of the 2003fg-like
supernova from the SD systems should be less than 0.8 Gyr while SN
2009dc may be from an old population, the model of Chen \& Li
(\cite{CHENWC09}) may thus account for a part of the 2003fg-like
supernovae.

In this paper, we found that the sub-luminous SNe Ia from
equal-mass DD system may at most account for 1 \% of all SNe Ia
observed. But Pakmor et al. (\cite{PAKMOR10}) found that their
model should occur with a rate of $\simeq2-11\%$ of the total SN
Ia rate, which seems consistent with the observation rate of
1991bg-like supernovae with errors ($16\%\pm7\%$, Li et al.
\cite{LIWD01}). However, their high birth rate is relative to
other possible SNe Ia progenitor formation channel, rather to that
from observations. Since no progenitor model may account for the
birth rate of SNe Ia derived observationally at
present\footnote{The DD model may explain the observational birth
rate of SNe Ia (Han \cite{HAN98}; Yungelson \& Livio \cite{YUN98,
YUN00}), but earlier numerical simulations showed that the most
probable fate of the coalescence of DD systems is an
accretion-induced collapse and, finally, neutron star formation
(Saio \& Nomoto \cite{SN85}; see also the review by Hillebrandt \&
Niemeyer \cite{HN00}).}, the birth rate in Pakmor et al.
(\cite{PAKMOR10}) might be overestimated. Then, the equal-mass DD
systems may be the progenitors of some1991bg-like supernovae.\\

\subsection{Age}\label{sect:4.2}
In the paper, we find that if an over-luminous SNe Ia is from a WD
+ MS system, its delay time should be shorter than 0.8 Gyr, which
means that there is new star formation during the recent 0.8 Gyr
in the host galaxies of the over-luminous SNe Ia if the SNe Ia are
from WD + MS systems. At present, there are four documented
examples of over-luminous SNe Ia explosions with an derived
progenitor mass higher than Chandrasekhar mass limit, i.e. SN
2003fg, 2006gz, 2007if and 2009dc. Amoung the four SNe Ia, only
the host galaxy of SN 2009dc was suspected to have no significant
information of star formation because it is an S0 galaxy and
inspection of its SDSS images show uniformly red color (Yamanaka
et al. \cite{YAMANAKA10}).

The delay time of 1991bg-like events is much different based on
the treatment of CE. For $\alpha=3.0$, the delay time is shorter
than 0.3 Gyr, while longer than 9 Gyr for $\alpha=1.0$. The range
of delay time for $\gamma$-algorithm is very wide, longer than
0.22 Gyr and even as long as 15 Gyr. Thus, our results provides a
possibility to test which treatment of CE may work for the CE
evolution of equal-mass DD systems via measuring the progenitor
age of 1991bg-like events. Now, many SNe Ia like SN 1991bg were
discovered and it is apparent that they favor E and S0 galaxies,
i.e. more than 60\% of 1991bg-like events are found in E/S0
galaxies and the remainder are found in early-type spiral galaxies
(Howell \cite{HOWELL01}), and then the sub-luminous SNe Ia are
expected to arise in old ($>1$ Gyr) stellar population (Sullivan
et al. \cite{SULLIVAN06}). For example, the host galaxy of SN
1991bg, NGC 4371, is an elliptical galaxy and its age is
approximately $14\pm2.5$ Gyr (Trager et al. \cite{TRAGER00}).
Although this age relies on the assumption of a single star burst
and is model dependent, the results of Trager et al.
(\cite{TRAGER00}) provided a lower limit of 10 Gyr for the age of
the progenitor of SN 1991bg (Howell \cite{HOWELL01}). The
simulation of $\alpha=1.0$ seems to favor such long age, but the
birth rate from the simulation is too low to compare with
observations, and the age range for the simulation is also too
narrow. Considering that the simulation of $\alpha=3.0$ also can
not explain the old population nature of 1991bg-like SNe Ia, the
$\alpha$-formalism may not explain the properties of these
sub-luminous events. In our simulation for $\gamma$-algorithm, the
Case 1 (older than 2 Gyr) dominates the sub-luminous SNe Ia and
contributes to more than 95\% of these SNe Ia. This means that
most of 1991bg-like events belong to old population, while there
is still a few the sub-luminous SNe Ia favoring young population,
which is more consistent with the statistic results of Howell
(\cite{HOWELL01}) than $\alpha$-formalism, at least qualitatively.
In addition, the birth rate for the $\gamma$-algorithm is also
more close to that from observations (Sullivan et al.
\cite{SULLIVAN06}). So, our results are more likely to favors the
$\gamma$-algorithm \footnote{Note: after analyzing the evolution
of a new sample of PCEBs, Zorotovic et al. (\cite{ZOROTOVIC10})
found that the classical $\alpha$-formalism seems to be an
appropriate prescription of CE evolution and turns out to
constrain the outcome of the CE evolution much more than the
$\gamma$-algorithm.}. One may argue that since the simulation of
$\alpha=1.0$ may produce the old sub-luminous SNe Ia as obtained
in Pakmor et al. (\cite{PAKMOR10}), my conclusion is possibly
dependent on the parameters of BPS simulation, such as the
distributions of mass ratio and eccentricity. We check the
influence of the distributions of mass ratio and eccentricity on
the birth rates of the special SNe Ia under the assumption of
$\alpha$-formalism. We find that the influence on the
super-luminous SNe Ia is insignificant (see Table \ref{Tab:1}),
while on the sub-luminous SNe Ia can not be neglected. The
distribution of $n(q)=2q$ favors the high mass ratio, and then may
enhance the contribution of Case 3 and the birth rate of the
sub-luminous SNe Ia, but the enhancement is moderate by about a
factor of 2 and the birth rate is still much lower than that
derived observationally (see Table \ref{Tab:1}). Elliptical orbit
is more possible to deduce the CE evolution than circular orbit.
So, under the distribution of $n(e)=2e$, the contribution of Case
1 is significantly enhanced, and then the birth rate of the
sub-luminous SNe Ia (see Table \ref{Tab:1}). However, the
distribution of $n(e)=2e$ almost dose not affect Case 3. Since the
delay time from Case 1 is very short, i.e. shorter than 0.3 Gyr,
most of 1991bg-like events (more than 97\%) should be from very
young population if $\alpha$-formalism works for the sub-luminous
SNe Ia, which conflicts with observations (Howell
\cite{HOWELL01}). So, our basic conclusion can not be influenced
by BPS parameters\footnote{In this paper, we do not test the
effect of mass-loss rate, instead of a classical Reimers¡¯s wind
(\cite{REI75}) with $\eta=0.25$ since the influence of mass loss
may be neglected when $\eta<1$ (Yang et al. \cite{YANGWM10}).}.
However, please keep in mind that our conclusion is critically
dependent on the result of Pakmor et al. (\cite{PAKMOR10}), i.e.
the 1991bg-like supernovae are from equal-mass DD systems, which
are only a part of SNe Ia from DD systems, not all DD systems.

In summary, we show the evolution of the birth rate of
over-luminous SNe Ia from SD systems and sub-luminous SNe Ia from
equal-mass DD systems in this paper, and found that the birth rate
of the SNe Ia are very low. Our results indicate that the model of
Chen \& Li (\cite{CHENWC09}) may account for some 2003fg-like
supernova, so as to the model of Pakmor et al. (\cite{PAKMOR10})
for 1991bg-like events. In addition, depending on the simulation
of Pakmor et al. (\cite{PAKMOR10}), the results in the paper
favors the $\gamma$-algorithm for the CE evolution of equal-mass
DD systems.

\begin{acknowledgements}
We are grateful to the anonymous referee for his/her constructive
suggestions which improve the manuscript greatly. This work was
supported by Natural Science Foundation of China under grant nos.
10873011, 10963001 and 11003003 and the Project of the Fundamental
and Frontier Research of Henan Province under grant no.
102300410223, and Program for Science \& Technology Innovation
Talents in Universities of Henan Province.
\end{acknowledgements}

\end{document}